\documentclass[aps,CPL,reprint,twocolumn,10pt,superscriptaddress,amssymb,amsmath,preprintnumbers,floatfix,showpacs,hyperref]{revtex4-1}

\usepackage[latin9]{inputenc}
\usepackage{bm}
\usepackage{color,CJK}
\usepackage{graphicx}
\usepackage{epsfig}
\usepackage[dvipdfmx,colorlinks,linkcolor=red,anchorcolor=blue,citecolor=blue]{hyperref}
\usepackage{upgreek}

\begin{document}

\title{Sub-Doppler laser cooling of $^{23}$Na in gray molasses on the $D_{2}$ line}

\renewcommand{\figurename}{Fig.}
\begin{CJK*}{GBK}{song}

\author{Zhenlian Shi}
\affiliation{State Key Laboratory of Quantum Optics and Quantum Optics Devices, \\  Institute of Opto-electronics, Shanxi University, Taiyuan, Shanxi 030006, China }
\affiliation{Collaborative Innovation Center of Extreme Optics, Shanxi University, Taiyuan, Shanxi 030006, China}

\author{Ziliang Li}
\affiliation{State Key Laboratory of Quantum Optics and Quantum Optics Devices, \\ Institute of Opto-electronics, Shanxi University, Taiyuan, Shanxi 030006, China }
\affiliation{Collaborative Innovation Center of Extreme Optics, Shanxi University, Taiyuan, Shanxi 030006, China}

\author{Pengjun Wang}
\email[Corresponding author email:]{pengjun$_$wang@sxu.edu.cn}
\affiliation{State Key Laboratory of Quantum Optics and Quantum Optics Devices, \\  Institute of Opto-electronics, Shanxi University, Taiyuan, Shanxi 030006, China }
\affiliation{Collaborative Innovation Center of Extreme Optics, Shanxi University, Taiyuan, Shanxi 030006, China}

\author{Zengming Meng}
\affiliation{State Key Laboratory of Quantum Optics and Quantum Optics Devices, \\ Institute of Opto-electronics, Shanxi University, Taiyuan, Shanxi 030006, China }
\affiliation{Collaborative Innovation Center of Extreme Optics, Shanxi University, Taiyuan, Shanxi 030006, China}

\author{Lianghui Huang}
\affiliation{State Key Laboratory of Quantum Optics and Quantum
Optics Devices, \\ Institute of Opto-electronics, Shanxi University,
Taiyuan, Shanxi 030006, China } \affiliation{Collaborative
Innovation Center of Extreme Optics, Shanxi University, Taiyuan,
Shanxi 030006, China}

\author{Jing Zhang}
\email[Corresponding author email:]{jzhang74@sxu.edu.cn; \\jzhang74@yahoo.com}
\affiliation{State Key Laboratory of Quantum Optics and Quantum Optics Devices, \\ Institute of Opto-electronics, Shanxi University, Taiyuan, Shanxi 030006, China }
\affiliation{Collaborative Innovation Center of Extreme Optics, Shanxi University, Taiyuan, Shanxi 030006, China}


\begin{abstract}
We report on the efficient gray molasses cooling of sodium atoms
using the $D_{2}$ optical transition at 589.1 nm. Thanks
to the hyperfine split about 6$\Gamma$ between the
$|F'=2\rangle$ and $|F'=3\rangle$ in the excited state
3$^{2}P_{3/2}$, this atomic transition is effective for the gray
molasses cooling mechanism. Using this cooling technique, the
atomic sample in $F = 2$ ground manifold is cooled from 700 $\upmu$K to 56
$\upmu$K in 3.5 ms. We observe that the loading efficiency into magnetic
trap is increased due to the lower temperature and high phase space
density of atomic cloud after gray molasses. This technique offers a
promising route for the fast cooling of the sodium atoms in the $F=2$ state.

\end{abstract}

\maketitle

Ultracold atomic gases offer a remarkably rich platform\cite{one1}
to enhance our understanding of numerous interesting phenomenon,
such as coherent manipulation of many body states and their dynamics using quantum gases in optical lattices,\cite{ol1,ol2,ol3,ol15} spin orbit
coupling to topological matter,\cite{ol4,ol5,ol6,ol7,ol8,ol9} and
ultracold association molecules to study the long range
dipole-dipole interaction and ultracold chemistry.\cite{one5,cpl7,one4}
New cooling techniques are developed continuously to produce an
atomic sample with a large atom number and lower temperature, such
as 3D Raman sideband cooling\cite{one7,image2,image4} and electromagnetically induced transparency (EIT) cooling\cite{image1,image3} in site-resolved imaging for Fermi atoms in optical lattice, and direct laser cooling of rubidium to quantum degeneracy in two-dimensional optical lattice.\cite{one9}

Cooling neutral atoms to ultracold temperature usually starts with a
magnetic-optical trap (MOT), and then transfers to optical trap or
magnetic trap for efficient evaporative cooling. Sub-Doppler laser
cooling is a powerful tool to decrease the temperature of a
three-level atomic system below the Doppler temperature
$\hbar\Gamma/2k_{B}$, where $\hbar$ and $k_{B}$ are the reduced
Plank constant and the Boltzmann constant, and $\Gamma$ is the
natural line width of atomic transition.\cite{two1,twoa1,twoa2}
This technique greatly increases the phase space density of atomic
cloud, resulting in the higher transfer efficiency to the trap and clearly leading to a gain in the final atom number after evaporative cooling.
Gray molasses (GM)
cooling is a special method for sub-Doppler laser cooling, which takes
advantage of polarization gradient cooling and velocity selective
coherent population trapping.\cite{two2}

In 1990s, GM cooling was proposed in Ref.\cite{three1}, in
which the fluorescence rate of the trapped atoms is strongly
reduced, and then was demonstrated experimentally on cesium\cite{three2,three3,three4} and rubidium\cite{three5} within the
$D_{2}$ transition. More recently, GM cooling was
realized on many atomic species, including alkali atoms $^{40}$K,\cite{three6,three7} $^{7}$Li,\cite{three8} $^{39}$K,\cite{three9,three10} $^{6}$Li,\cite{three7,three11} $^{23}$Na,\cite{three13} $^{41}$K,\cite{three41k} and metastable atom
$^{4}$He,\cite{three12} which was operating on the blue detuning of
the $D_{1}$ transition with more resolved energy spectrum. The
cooling technique was also implemented on the $D_{2}$ transition to
cool $^{40}$K to 50 $\upmu$K with red-detuned laser,\cite{three14}
and cool $^{87}$Rb to 4 $\upmu$K with blue-detuned laser.\cite{three15} It has been proven experimentally that the GM cooling leads to substantial advantages in terms of lower temperature and higher phase-space density. In the GM
cooling mechanism, dark and bright states are the coherent
superposition of Zeeman sublevels in the ground hyperfine manifold.\cite{three7} The bright state energy is spatially modulated, giving
a polarization gradient cooling like a standard red detuned
molasses. Atoms with higher energy in dark states could be
coupled to the bright states by motion induced coupling and then could be
cooled down in the polarization gradient cooling, and ones with
lower energy are trapped in dark space which prevent heating induced
by light-assisted collisions.

In this Letter, we report on an experimental study of sodium atoms
cooled in six beams GM on the blue side of the $D_{2}$
transition $ |F_{g}=2\rangle\rightarrow|F_{e}=2\rangle$, as shown in
Fig.1(a). Thanks to the much reduced fluorescence rate in GM compared to the bright molasses cooling, we are
able to capture cold and dense atomic cloud of $ 3\times 10^{8} $ atoms
at temperature of 56 $\upmu$K. After the GM phase, atoms
in $ |F=2,m_{F}=2\rangle$ are transferred to an optically plugged
quadrupole trap, where $F$ is the total angular momentum and $m_{F}$
is its projection. We observe that the loading efficiency is increased
to 70\% due to the lower temperature and high phase space
density of atomic cloud using the GM technique. We also
show a significant evidence of the atom's temperature and
number after a short RF-induced evaporative cooling as a function of
the two-photon Raman detuning in GM phase.

\begin{figure}
\centerline{\includegraphics[width=3.7in]{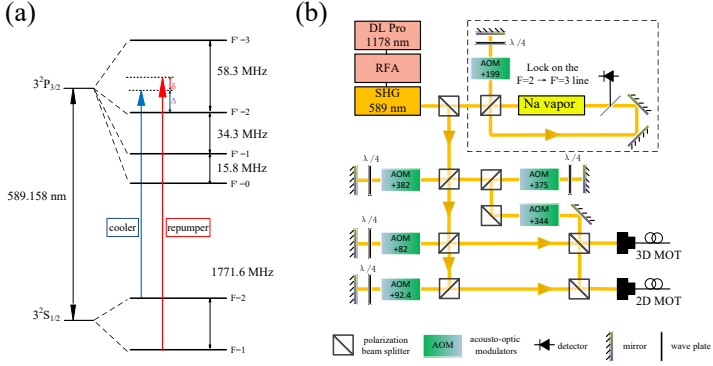}}

\caption{(a) Sodium cooling scheme on the $D_{2}$ line. The cooling beam (blue) is blue detuned by $\varDelta$  from $|F_{g}=2\rangle\rightarrow|F_{e}=2\rangle$ transition, the repumping beam (red) is blue detuned by $\varDelta$ + $\delta$ from $ |F_{g}=1\rangle\rightarrow|F_{e}=2\rangle$ transition, where the $F_{g}(F_{e})$ is the ground-excite state level. Here, $\delta$ denotes the two-photon Raman detuning. (b) Sketch of the optical setup for cooling sodium, and the dashed box represents the setup for saturated absorption spectroscopy.  \label{Fig1} }

\end{figure}

In the experiment, the laser light (at 589 nm) is produced by
frequency doubling a master oscillator power amplifier (MOPA)
system. First, the laser beam (at 1178 nm) of 23 mW from external cavity
diode laser (ECDL) is seeded to a Raman fiber amplifier (RFA), to be
amplified to 7 W and transferred to the SHG unit for second harmonic
generation (SHG) modules. We obtain 3.3 W of output power at 589 nm,
which is sufficient for cooling and trapping of $^{23}$Na in
experiment. This single laser source supplies the coherence between
the two frequencies for generating the long-lived dressed dark
states in GM cooling. The frequency of output beam is
stabilized at $ |F_{g}=2\rangle\rightarrow|F_{e}=3\rangle$ by the
saturated absorption spectroscopy. The cooling and repumping laser
beams are obtained using acousto-optic modulators (AOMs) in
double pass configuration.\cite{doublepass} Here, the AOMs are
acting as fast switches and frequency and intensity tuners. After
that the laser beams of the cooling and repumping laser are combined by a polarization beam splitter (PBS)
and delivered to the science cell by polarization maintaining
optical fibers; as shown in Fig.\ref{Fig1}(b). They are then expanded to 25 mm for MOT beams and
distributed to the three pairs of $\sigma^{+}$ - $\sigma^{-}$
counter-propagating configuration.

Sodium atoms are loaded from a two dimensional magnetic-optical trap and trapped by a dual-operation three dimensional magnetic-optical trap (3DMOT) operating on the $D_{2}$ transition introduced in,\cite{mot1, mot2} in which two laser frequencies are used, one is tuned to $ |F_{g}=2\rangle\rightarrow|F_{e}=3\rangle$ transition with a red detuning 34 MHz for cooling and the other one is tuned to $ |F_{g}=1\rangle\rightarrow|F_{e}=2\rangle$ transition with a red detuning 57 MHz for repumping. In the 3DMOT, the almost same number of atoms in the $F$ = 1 and $F$ = 2 states and total number is about $\ 7.3\times 10^{9}$. The temperature is about 700 $\upmu$K. At the end of 3DMOT loading, the compressed magnetic optical trap (CMOT) is used to increase the density, in which the magnetic field gradient is ramped from 4 G/cm to 10 G/cm in 50 ms.

To decrease the temperature, the GM phase is applied for a duration time 3.5 ms after the magnetic field is switched off in $\simeq$ 100 $\upmu$s. Here, we define the detuning of the cooling frequency from the $ |F_{g}=2\rangle\rightarrow|F_{e}=2\rangle$ transition as the principal (one-photon) laser detuning $\varDelta$ in the range of $1\Gamma - 5\Gamma$ ($\Gamma=2\pi\times9.79$ MHz is the natural linewidth of the $D_{2}$ line), and the frequency difference between the cooler $f_{C}$ and the repumper $f_{R}$ as the two-photon Raman detuning $\delta=f_{R}-f_{C}-1771.6$ MHz. During the molasses time t, the cooling and repumping beams are detuned to desired values and the intensity of two laser beams is reduced. At the end of GM phase, an absorption image is applied to measure the temperature and number of atomic cloud after a few ms of time of flight (TOF).

\begin{figure}
\centerline{
\includegraphics[width=3.2in]{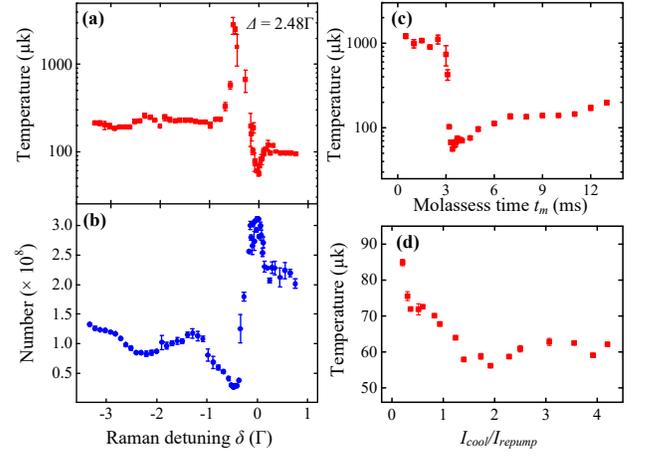}
} \vspace{0.1in}

\caption{(a) The temperature (red squares) and (b) number (blue circles) of atomic sample after GM with time $t$ = 3.5 ms as a function of two photon Raman detuning in the case of $\varDelta$ = 2.48$\Gamma$.
The temperature after GM at the fix principal detuning $\varDelta$ = 2.48$\Gamma$ and Raman detuning $\delta$ = 0, as a function of (c) the pulse duration time $t$ and (d) the intensity ratio $I_{cool}/I_{repump}$ of the Raman cooling pulse at a constant intensity $I_{repump}=5I_{sat}$, where $I_{sat}=6.26$ $\text{mw/cm}^{2}$. The data points show the average of three experimental measurements, with error bars corresponding to the standard deviation of the mean.
\label{Fig2} }

\end{figure}

We first study the effect of GM cooling for different experimental parameters with the fixed principal detuning $\varDelta$ = 2.48$\Gamma$ (red detuned 34 MHz on the transition of $ |F_{g}=2\rangle\rightarrow|F_{e}=3\rangle$, which means that it is the blue detuning 24.3 MHz for $ |F_{g}=2\rangle\rightarrow|F_{e}=2\rangle$), such as two-photon Raman detuning, duration time of GM and the intensity ratio, as shown in Fig.\ref{Fig2}.

The atomic sample temperature and number after the GM are observed as a function of two-photon detuning $\delta$ and a Fano-like profile at the atom temperature due to the interference of different excitation processes, accompanied by a sharp change in the number of cooled atoms, as shown in Figs.\,2(a) and 2(b). Firstly, we focus on the case of the zero Raman detuning $\delta=0$, where we find that the temperature minimum of $\emph{T}=$ 56 $\upmu$K is reached. The maximum number about $\ 3.1\times 10^{8}$ of the atomic cloud is obtained by the GM with a phase space density (PSD) of $\ 1.3\times 10^{-6}$, which means that in the case the atoms in dark spaces are maximally coupled to bright states at the bottom of energy hills and this induces a peak in the atom number in Fig.\ref{Fig2}(b). The motion coupling between dark states and bright states is velocity selective, thus the atoms with higher velocity could be coupled to bright states and re-cooled by the Sisyphus cooling, and the coldest atoms will accumulate in the dark spaces with less interaction with the light field. The GM cooling mechanism is working in this case and could cool the atoms to lower temperature than the bright molasses on the $D_{2}$ line. For slightly red detuned at $\delta=-0.5\Gamma$, a heating peak is observed, which means that in the case the atoms in dark states have more energy than those in bright states and will gain kinetic energy when they couple to bright states at the top of energy hills. This mechanism leads a strong heating, and the temperature increases from 700 $\upmu$K (the MOT temperature) to about 2 mK, accompanied by a significant loss of atom number during the molasses phase, which is the inverse of the Sisyphus cooling. In addition to the sharp dip and peak in atom temperature and number close the resonance $\delta$ = 0, we notice that the more atoms with lower temperature are captured in the case of $\delta > 0$ compared with that at the red detuning $\delta < 0$. This phenomenon is corrected that the main component of the dark state is lower than that of the bright state in the detuning $\delta > 0$, and more atoms could be trapped in dark states with lower temperature.

\begin{figure}
\centerline{
\includegraphics[width=3.3in]{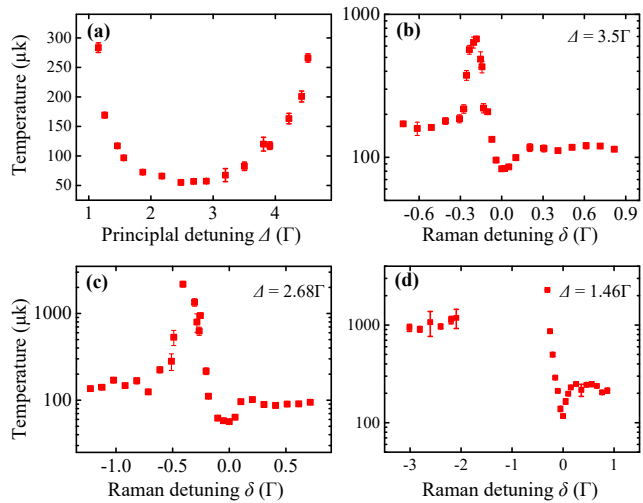}
} \vspace{0.1in}
\caption{(a) The minimum temperature obtained after GM as a function of the principal detuning $\varDelta$; The atom temperature of $^{23}$Na after $t$ = 3.5 ms of GM cooling as a function of two photon Raman detuning in the case of (b) $\varDelta$ = 3.5$\Gamma$, (c) $\varDelta$ = 2.68$\Gamma$ and (d) $\varDelta$ = 1.46$\Gamma$.
\label{Fig3} }

\end{figure}

The temperature after the GM is plotted as a function of the duration time $t_{m}$ in Fig.\ref{Fig2}(c). The temperature decreases immediately at about 3 ms and reaches the minimum value at about 3.5 ms where a steady state is built, and then increases slowly with a longer duration time. The observation of no noticeable cooling effect at first 3 ms is attributed to the small misplacement between the center of magnetic trap and the position of atoms cloud, which is designed for better loading in the MOT phase to avoid atoms loss. Figure 2(d) shows the temperature as a function of the ratio between cooler and repumper intensities $I_{cool}/I_{repump}$. We find a nearly linearly decrease at the initial regimes and a minimal temperature of $\sim$ 56 $\upmu$K reached at the ratio of 2, and a slightly increase with more cooling power. This may be understood from the induced light shift in the dressed state picture.\cite{three8} At the case of fixed frequency, the atoms remained in dark states distribute the spatial variation of the standing wave and loss how much energy depending on the difference of light shift between the entry and departure points, as in the behavior of bright molasses.

We also measure the atom temperature after the GM with different one-photon detunings for $1\Gamma<\varDelta<5\Gamma$ on the blue side of the transition of $ |F_{g}=2\rangle\rightarrow|F_{e}=2\rangle$, as shown in Fig.\ref{Fig3}. This shows that the minimal temperature could reach $\sim$ 55 $\upmu$K for $\varDelta$ = 2.48$\Gamma$, and increases on both sides. This behavior can be attributed to the chosen structure in GM: the energy separation between the $|F_{e}=2\rangle$ and $|F_{e}=3\rangle$ in excite states is about $5.9\Gamma$, thus for higher $\varDelta$, the upper state may start to play a role. As mentioned in Ref.\cite{three15}, the closed transition of $ |F_{g}=2\rangle\rightarrow|F_{e}=3\rangle$ contributes to the inefficiency of the GM cooling. This phenomenon is different from that of GM on the $D_{1}$ optical transition, in which the reached temperature is weakly dependent on the one-photon laser detuning.

The measured temperature after GM with different one-photon detunings as a function of the two photon Raman detunings $\delta$ are shown in Figs.\,3(b)--3 (d). For each value of $\varDelta$, the minimum temperature is obtained on $\delta\approx0$, similar to the report in earlier experiments. In the case of $\varDelta$ = 1.46$\Gamma$, there is a very wide heating window and the temperature is very higher at the red detuning, where the repumper is close resonance with the atomic transition of $ |F_{g}=1\rangle\rightarrow|F_{e}=2\rangle$, as shown in Fig.\ref{Fig3}(d). During optimizing the GM, we find that the stray magnetic field and misalignment of the laser beams affects the attained minimal temperature of the GM phase. Here, the three pairs of coils are used to produce $XYZ$ bias fields for compensating the stray field.

\begin{figure}
\centerline{\includegraphics[width=3.2in]{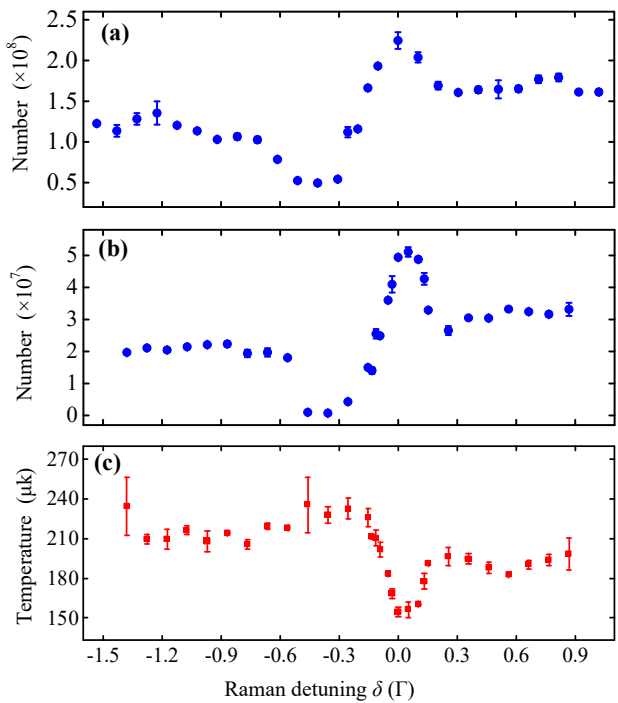}}
\caption{(a) The atom number loaded into the magnetic trap as a function of the two-photon Raman detuning.
The number (b) and temperature (c) of atoms after a short RF forced evaporative cooling as a function of the two-photon Raman detuning in GM phase at $\varDelta$ = 2.48$\Gamma$.
\label{Fig4} }
\end{figure}

The most evident advantage of the GM is to obtain a large number of atoms with lower temperature for loading to magnetic trap. After the GM phase, optical pumping is applied in 0.8 ms, during which the atoms are pumped to the stretched low field seeking Zeeman state $|F=2,m_{F}=2\rangle$, and then are loaded to the optically plugged magnetic trap. In the loading magnetic trap process, a magnetic trap with lower gradient about 10 G/cm is fist used and then the trap gradient is increased to 200 G/cm in 200 ms. The special designed ramp stage can prevent the atoms in other low field seeking hyperfine states from loading to the magnetic trap. Lastly, atoms loaded in magnetic trap can be probed after a few ms of time-of-flight (TOF). The atom number in the magnetic trap as a function of the two-photon Raman detuning in GM is shown in Fig.\ref{Fig4}(a). Here the one-photon detuning is $\varDelta$ = 2.48$\Gamma$. We also observe a Fano-like profile, from which the loading efficiency is increased to 70\% by the GM cooling mechanism.

After loading to the magnetic trap, an RF forced evaporative cooling is performed to cool the atomic sample down to lower temperature for 5\,s. We also study the effect of gray molasses on this stage. The atom number and temperature after evaporative cooling versus the two-photon Raman detuning in gray molasses are shown in Figs.\,4(b) and 4(c). After optimizing the parameters on the gray molasses, an atomic sample containing $5.1\times10^{7}$ atoms with $\emph{T}$ = 150 $\upmu$K is produced.

In conclusion, we have shown an effective GM cooling of sodium on the $D_{2}$ transition,
thanks to the sufficiently large hyperfine split about $6\Gamma$ between the $|F_{e}=2\rangle$ and $|F_{e}=3\rangle$ in the excited state 3$^{2}P_{3/2}$, and the phase coherence between the cooling and repumping beams from a single laser source.
We have investigated the properties of GM cooling by studying the dependence of the sample temperature on the one-photon detuning, two-photon Raman detuning, molasses duration time and the intensity ratio.
The high phase space density after GM provides a starting condition for the effective loading of magnetic traps, which is the best evidence of the cooling mechanism. The experimental results give us a picture of GM for $D_{2}$ line which could work in the atomic species with the hyperfine split about $\sim6\Gamma$ in excited states. Our experiment shows that GM cooling on the $D_{2}$ line is a viable route to high phase space density, while it is lower than in GM cooling on $D_{1}$ line. The requirement of only a single laser for all cooling process is a clear advantage.

\begin{acknowledgments}

This research was supported by National Key Research and Development Program of China (Grant No. 2016YFA0301602),
NSFC (Grants No. 11474188, and No. 11704234),
the Fund for Shanxi ``1331 Project'' Key Subjects Construction,
and the program of Youth Sanjin Scholar.

\end{acknowledgments}

\end{CJK*}

\end{document}